\documentclass[aps,prb,twocolumn,amsmath,amssymb,showpacs,floatfix]{revtex4-1}
\usepackage{hyperref}
\usepackage{graphicx}
\usepackage{dcolumn}
\usepackage{bm}
\usepackage[latin1]{inputenc} 
\usepackage{color} 
\begin{document}

\title{Tunneling spectroscopy probing magnetic and nonmagnetic electrodes in tunnel junctions}
\author{Volker Drewello}
\email{drewello@physik.uni-bielefeld.de}
\author{Zo\"e Kugler}
\author{G\"unter Reiss}
\author{Andy Thomas}
\affiliation{Bielefeld University, Thin Films and Physics of Nanostructures, 33615 Bielefeld, Germany}
\date{\today}
\begin{abstract}
Tunneling spectroscopy is applied to tunnel junctions with only one or no ferromagnetic electrode to study the excitation of quasi particles in magnetic tunnel junctions. The bias dependence is investigated with high accuracy by inelastic electron tunneling spectroscopy. Both types of junctions show a zero bias anomaly that is different in size and sign compared to magnetic tunnel junctions, i.e. junctions with two ferromagnetic electrodes. A pronounced difference is also found depending on the material that is probed by the tunneling electrons, which might be attributed to the excitation of magnons.
\end{abstract}

\maketitle

Magnetic tunnel junctions (MTJs) with MgO as a crystalline barrier show large tunnel magnetoresistance (TMR) ratios of up to 1000\,\% at low temperature \cite{Ikeda2008,Tezuka2009}. Nevertheless, the TMR ratio at room temperature is still by a factor of 2 to 3 smaller. Decreasing this temperature dependence is one way to gain higher TMR ratios at room temperature and increase the applicability of MTJ based spintronic devices. The main reason for the decreasing TMR values are intrinsic excitations within the junctions \cite{Bratkovsky1998,Han2001, Drewello2008} which can be studied by means of inelastic electron tunneling spectroscopy (IETS). In this paper, the different excitations in magnetic and nonmagnetic electrodes are investigated. Tungsten is chosen as the nonmagnetic metal electrode because of its high melting point. It also has a lower lattice mismatch with the MgO barrier than e.g. Tantalum (6\,\% vs. 11\,\%).

The samples are prepared in a self-made sputter deposition tool with a base pressure below $10^{-9}$\,mbar. The layers are deposited on top of thermally oxidized silicon wafers at an Ar pressure of $5\times10^{-3}$\,mbar (MgO at $10^{-2}$\,mbar).  The first sample is a metal/ insulator/ metal tunnel junction (M-I-M)  with the layer stacking W 20/ MgO 1.8/ W 20 (all values in nm). The second sample has a ferromagnet as the lower electrode (FM-I-M). The layer stack is W 15/ Co$_{40}$Fe$_{40}$B$_{20}$ 6/ MgO 1.8/ W 20. Both stacks are annealed at 723\,K for 1 hour in a high vacuum furnace. A capping of Ta and Au is added for protection and to form contact pads. The samples are structured by optical lithography and Ar ion beam etching. 

The low temperature measurements are done at 13\,K in a closed cycle Helium cryostat by a standard two-probe technique. The bias voltage is defined with respect to the upper electrode. Thus, negative bias results in electrons tunneling into the upper electrode.
A Lock-In technique is used to obtain the d$I$/d$V$ curves which are differentiated numerically to get the IET spectra (d$^2I/$d$V^2$). 
Details of the measurement setup and procedure can be found elsewhere \cite{Drewello2009}.

\begin{figure}[b]
	\centering
	\includegraphics[scale=1]{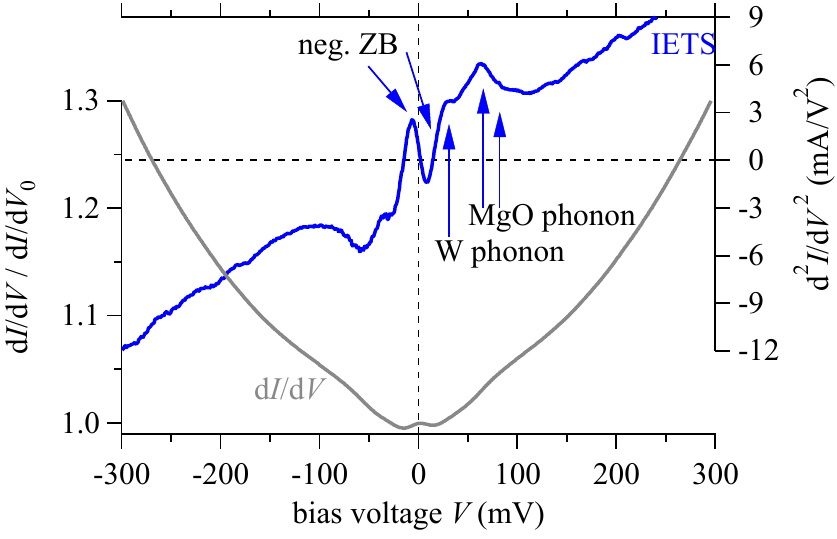}
 	 \caption{The tunnel spectrum (d$I$/d$V$) normalized to its value at $V=0$ and the IET spectrum (d$^2I$/d$V^2$) of the W/ MgO/ W sample.}
	\label{fig1}
\end{figure}

\begin{figure}[tb]
	\centering
	\includegraphics[scale=1]{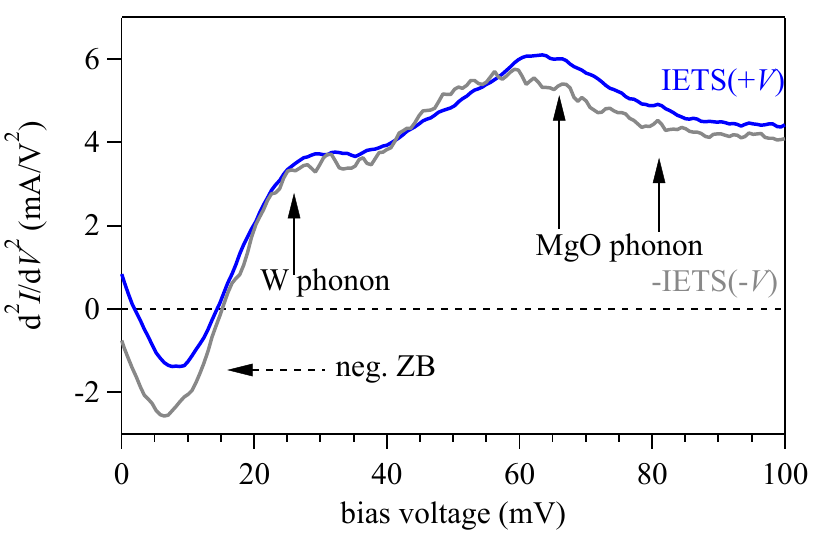}
 	 \caption{The IET spectrum of the W/ MgO/ W junction. The low bias region is shown for both polarities. The arrows mark energies of known excitations.}
	\label{fig2}
\end{figure}

The spectra for the M-I-M junction are shown in Figure \ref{fig1}. The conductance has the predicted parabolic shape \cite{Simmons1964} with variations at bias voltages up to 100\,mV. In the IET spectrum these features can be seen much clearer. First, at approximately 9\,mV a negative zero bias anomaly is found, i.e. the conductance decreases at low bias compared to zero bias. In typical full MTJs (i.e. with two magnetic electrodes) the zero bias anomaly shows the opposite sign. Often the magnitude of the effect is larger and dominates the spectra \cite{Moodera1998,Miao2006,Drewello2009}.
Second, several broad peaks are found up to 100\,mV. Figure \ref{fig2} shows these features in more detail. As no magnetic materials are used and no magnetic impurities are expected, phonon excitation is an explanation for these peaks. Peaks of the electrode phonons typically have energies around 30\,meV \cite{Klein1973} while the phonons of barrier oxides have higher energies. Here, the first peaks correspond to an excitation of tungsten phonons with an energy of 26\,meV \cite{Olejniczak1998}. 
The MgO tunneling barrier leads to phonon peaks at 66 and 81\,mV \cite{Klein1973}. 
A strong peak is indeed found at 66\,mV but only a shoulder is found at 81\,mV. In MgO based full MTJs this peak is typically more pronounced \cite{Miao2006, Drewello2009}.

\begin{figure}[tb]
	\centering
	\includegraphics[scale=1]{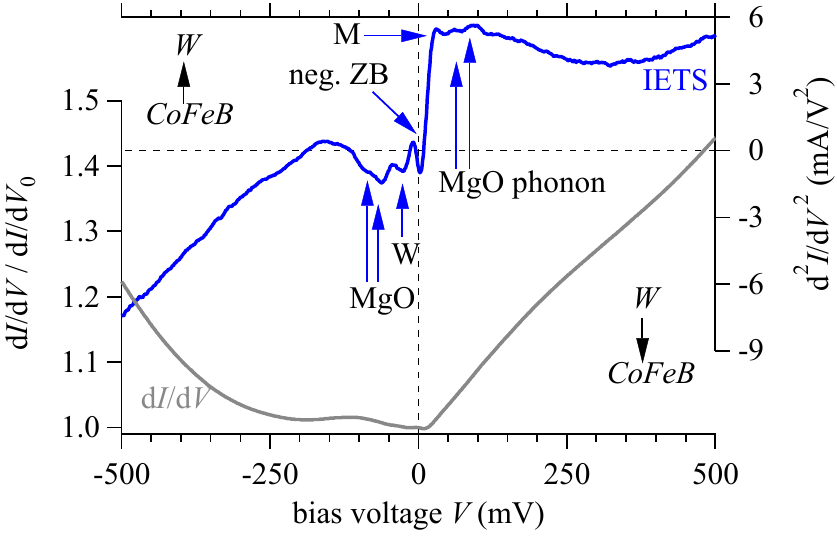}
  	\caption{The normalized tunnel spectrum and the IET spectrum of the Co-Fe-B/ MgO/ W sample. }
	\label{fig3}
\end{figure}

\begin{figure}[tb]
	\centering
	\includegraphics[scale=1]{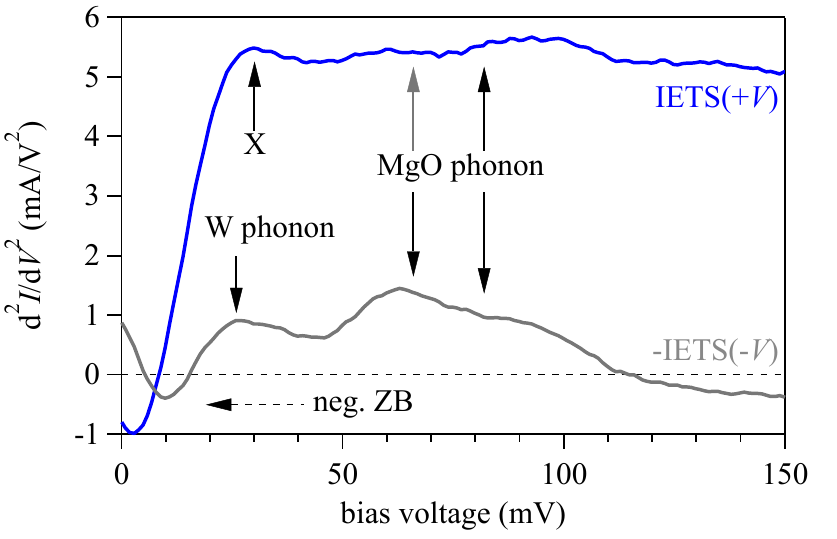}
  	\caption{Comparison of the IET spectra for positive (electrons tunnel into Co-Fe-B) and negative (into W) bias voltage for the Co-Fe-B/ MgO/ W sample. }
	\label{fig4}
\end{figure}

The spectra of the FM-I-M junction show a strong asymmetry in the slope of the dI/dV-curves (Figure \ref{fig3}). This leads to a different height of the IET spectra for positive compared to negative bias, nevertheless the peak structure is visible. Magnified spectra for negative (electrons tunnel into W) and positive bias (electrons tunnel into Co-Fe-B) can be compared in Figure \ref{fig4}. Most prominent is the zero bias feature which is still negative for both bias polarities. For negative bias it is found at approximately 10\,mV, while it is closer to zero for positive bias. This is presumably caused by smearing with the following peaks, which have a much higher intensity (X). For both polarities several features are found. The spectrum for negative bias looks similar to the spectrum of the first sample, with the tungsten phonon peak at 26\,mV and a MgO phonon peak at 66\,mV. The 81\,mV shoulder is more pronounced. For positive bias, i.e. tunneling into the ferromagnet, the peaks are higher and less sharp. After the neagtive ZBA the spectrum rises strongly to the first peak. This must be the excitation of magnons (X). Also, the MgO phonon peak at 81\,mV peak is more pronounced than the one at 66\,mV.

\begin{figure}[tbp]
	\centering
	\includegraphics[scale=1]{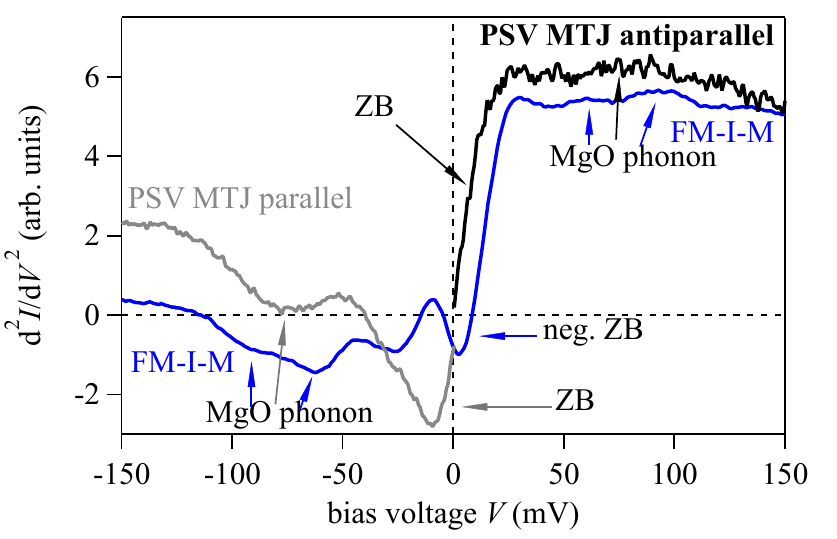}
  	\caption{Comparison of the IET spectra of the Co-Fe-B/ MgO/ W sample and those of a typical MTJ (scaled). Arrows mark the ZBA.}
	\label{fig5}
\end{figure}

In Figure \ref{fig5} the IET spectra are compared to those of a typical MTJ \footnote{ The full MTJ is a pseudo spin valve with a similar layer stack: Ta 20 / Co-Fe-B 5.4/ MgO 2.4/ Co-Fe-B 2.4/ Ta 20. It is annealed at 723\,K for 1 hour.}.The FM-I-M spectrum for tunneling into W (negative bias) is compared to the spectrum for the parallel (P) magnetic state. For tunneling into the FM (pos. bias) the spectrum is compared to the MTJ in antiparallel (AP) magnetic state.
 In both cases the spectra have a similar shape. The striking difference is the zero bias anomaly, which is positive for the MTJ and leads to a huge peak in the P state. This effect is also visible in the AP state, but the zero bias peak is only a shoulder. As a result the spectrum immediately rises after zero bias, while for the FM-I-M junction the flank seems shifted to higher bias. The MgO phonon peaks are also similar in both cases. The spectra of the full MTJ  in the P state are negative over a wide bias range. In the AP state, the spectra are much lower compared to the FM-I-M-spectra in the same bias region. 

Now the results will be discussed. In the presented spectra most peaks can be identified as belonging to the target electrode and the barrier with the exception of the zero bias anomaly (ZBA). Its size and sign is obviously not simply a matter of the material of the target electrode. This can be seen from the results of the FM-I-M junction, where the ZBA is roughly the same for the two different electrode materials. Furthermore, the same ferromagnet shows large positive ZBA peaks in the MTJ. The two different results suggest that it is not a excitation of surface magnons which is causing the ZBA.
This leaves tunneling through impurities as an explanation. It is known that the zero bias effect depends on the impurity material \cite{Cooper1973}.  
A model that would explain impurities in the barrier is implantation of upper electrode atoms during preparation. In this case the material would be tungsten in both samples (in difference to the MTJ case). 

\begin{figure}[tbp]
	\centering
	\includegraphics[scale=1]{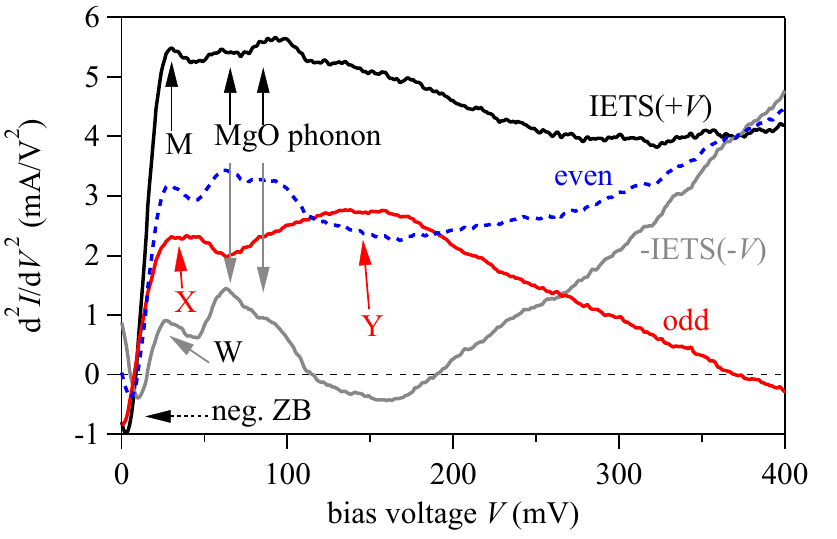}
 	\caption{Comparison of IET spectra for positive/negative bias and even and odd spectra of the Co-Fe-B/ MgO/ W sample.}
	\label{fig6}
\end{figure}

The M-I-M sample shows near ideal symmetry and the anticipated peaks. The asymmetry in the spectra of the FM-I-M sample, however, rises the question of magnon excitation in the ferromagnet. If even/odd spectra \footnote{The even (odd) spectrum is the average (difference) of the spectra for positive and negative bias.}  are calculated, peak structures are indeed found in the odd spectrum (Figure \ref{fig6}). The first peak must be the excitation of magnons, as observed before. However, the origin of the second, much broader peak is not clear. It's maximum is at high bias voltage around 150\,mV and leads up to some hundred mV. This result is different compared to the peak at 10\,mV that Paluskar et al. find for incoherent Alumina based junctions \cite{Paluskar2007}. The high energies we find would be equivalent to temperatures higher than 1000 K and, therefore, $T_C$. The total magnon density of states in a ferromagnet is large at these energies \cite{Halilov1997} and bulk magnons are suspected to contribute to the tunneling process at some voltage \cite{Bratkovsky1998}. However, it is not clear if the interaction potential allows these modes to be excited \cite{Balashov2008}.

Nevertheless, the similarity of the FM-I-M spectra to the state-specific spectra of the MTJ also fits in this model. In the case of electrons tunneling into Co-Fe-B magnons should be the dominant excitation. 
This is also the case in the AP state of a MTJ where the direct tunneling contribution is smaller due to the inverse spinpolarization of the electrodes. As the major difference is caused by the anomaly around zero bias, it can be suspected that the broad contribution, which is not seen in the P state, is also the excitation of magnons. 
Lastly, the broad dip (or gap) in the tunnel spectrum (d$I$/d$V$) of the FM-I-M junction resembles the P state spectrum of MTJs that incorporate one half-metallic Heusler compound electrode \cite{Sakuraba2006, DrewelloAPL2009}. In both cases magnon excitation is prohibited for the corresponding bias polarity, while it is allowed for the other one. This is another hint that the magnon excitation is the origin of the broader feature (gap in d$I$/d$V$, background in IETS). 

A minor note regards the MTJ spectrum in the P state. It is shifted to negative values, which might indicate a large coherent tunneling contribution \cite{Tsunegi2008}. This is not seen in the M-I-M sample, which might be tentatively ascribed to different growth of the MgO barrier depending on the lower electrode material. The different strength of the MgO phonons peaks could then also be a effect of different barrier and interface properties. 

In summary, tunnel junctions with only one or no ferromagnetic electrode have been investigated by inelastic electron tunneling spectroscopy. The excitations of electrode and barrier phonons are observed in all junctions. For the junction with one ferromagnetic electrode the excitation spectra show a strong asymmetry, which is attributed to magnon excitation. In contrast to full magnetic tunnel junction, the presented junctions show a negative zero bias anomaly.

\begin{acknowledgments}
We gratefully acknowledge 
Jan Schmalhorst for helpful discussions, 
Patryk Krzysteczko and Markus Sch\"afers for technical assistance, and 
the DFG (Grant RE 1052/13-1) for financial support. 
\end{acknowledgments}

\bibliography{paper-database}

\end{document}